\documentclass[traditabstract]{aa}
\usepackage{tabularx}
\usepackage{amssymb}
\usepackage{amsmath}
\usepackage{txfonts}
\usepackage{balance}
\usepackage{natbib}
\bibpunct{(}{)}{;}{a}{}{,} 
\usepackage{color}
\usepackage{xspace}
\usepackage[normalem]{ulem} 
\usepackage[table]{xcolor} 
\usepackage{multirow}
\usepackage{rotating}
\usepackage{bbm}
\usepackage{cancel}
\usepackage{graphicx}
\usepackage{caption}
\usepackage{ifthen}
\usepackage{booktabs}
\usepackage{tabularx}
\usepackage{float}
\usepackage{gensymb}

\makeatletter
\if@referee
    \newcommand{\com}[1]{\textcolor{red}{[#1]}}                            
    \usepackage[nolists,noheads,nomarkers]{endfloat}                       
\else
    \newcommand{\com}[1]{}                                                 
\fi
\makeatother
\usepackage[%
pdfauthor={Fredrik Windmarkl},
pdftitle={Velocity distribution},
pdfstartview=FitH,
linkcolor=blue,
anchorcolor=blue,
citecolor=blue,
filecolor=blue,
menucolor=blue,
urlcolor=blue,
colorlinks=true]{hyperref}
 
\begin{document}
\title{Passive galaxies as tracers of cluster environments at
  $z\sim2$} \titlerunning{Passive overdensities at $z\sim2$}
\authorrunning{Strazzullo et al.}  \author{V. Strazzullo\inst{1,2},
  E. Daddi\inst{1}, R. Gobat\inst{1}, B. Garilli\inst{3},
  M. Mignoli\inst{4}, F. Valentino\inst{1}, M. Onodera\inst{5},
  A. Renzini\inst{6}, A. Cimatti\inst{7}, A. Finoguenov\inst{8},
  N. Arimoto\inst{9}, M. Cappellari\inst{10}, C.~M. Carollo\inst{5},
  C. Feruglio\inst{11}, E. Le Floc'h\inst{1}, S.~J. Lilly\inst{5},\\ D. Maccagni\inst{3}, H.~J. McCracken\inst{12}
  M. Moresco\inst{7}, L. Pozzetti\inst{4}, G. Zamorani\inst{4} }

\institute{Irfu/Service d'Astrophysique, CEA Saclay, Orme des
  Merisiers, F-91191 Gif sur Yvette, France -- e-mail:
  vstrazz@usm.lmu.de \and Department of Physics,
  Ludwig-Maximilians-Universit{\"a}t, Scheinerstr. 1, 81679
  M{\"u}nchen, Germany \and INAF -- IASF, via Bassini 15, I-20133,
  Milano, Italy \and INAF - Osservatorio Astronomico di Bologna, via
  Ranzani 1, 40127 Bologna, Italy \and Institute for Astronomy, ETH
  Z{\"u}rich, Wolfgang-Pauli-strasse 27, 8093 Z{\"u}rich, Switzerland
  \and INAF-Osservatorio Astronomico di Padova, Vicolo
  dell'Osservatorio 5, I-35122, Padova, Italy \and Dipartimento di
  Fisica e Astronomia, Universit{\'a} di Bologna, Viale Berti Pichat
  6/2, I-30127, Bologna, Italy \and Department of Physics, University
  of Helsinki, Gustaf H{\"a}llstr{\"o}min katu 2a, FI-0014 Helsinki,
  Finland \and National Astronomical Observatory of Japan, Subaru
  Telescope, 650 North Aohoku Place, Hilo, HI 96720, USA \and
  Department of Physics, University of Oxford, Denys Wilkinson
  Building, Keble Road, Oxford, OX1 3RH, UK \and IRAM - Institut de
  Radioastronomie Millim{\'e}trique, 300 rue de la Piscine, 38406
  Saint Martin d’H{\`e}res, France \and Institut d’Astrophysique de
  Paris, UMR7095 CNRS, Universit{\'e} Pierre et Marie Curie, Paris,
  France}

\date{ } 

\abstract {Even 10 billion years ago, the cores of the first galaxy
  clusters are often found to host a characteristic population of
  massive galaxies with already suppressed star formation. Here we
  search for distant cluster candidates at $z$$\sim$2 using massive
  passive galaxies as tracers. With a sample of $\sim$40
  spectroscopically confirmed passive galaxies at 1.3$<$$z$$<$2.1, we
  tune photometric redshifts of several thousands passive sources in
  the full 2~sq.~deg.  COSMOS field. This allows us to map their
  density in redshift slices, probing the large scale structure in the
  COSMOS field as traced by passive sources.  We report here on the
  three strongest passive galaxy overdensities that we identify in the
  redshift range 1.5$<$$z$$<$2.5. While the actual nature of these
  concentrations is still to be confirmed, we discuss their
  identification procedure, and the arguments supporting them as
  candidate galaxy clusters (likely mid-10$^{13}$M$_{\odot}$
  range). Although this search approach is likely biased towards more
  evolved structures, it has the potential to select still rare,
  cluster-like environments close to their epoch of first appearance,
  enabling new investigations of the evolution of galaxies in the
  context of structure growth. }

 
\maketitle

\section{Introduction}
\label{sec:intro}

Up to at least $z$$\sim$1, passive galaxies typically with early-type
morphology dominate the high-mass end of the galaxy population, and
are the best tracers of the highest density peaks in the large scale
structure.  The evolution of passive galaxy populations at
$z$$\lesssim$1 -- and in particular with respect to environmental
effects -- has been explored in detail also thanks to large
spectroscopic campaigns
\citep[e.g.,][]{kauffmann2004,bernardi2006,gallazzi2006,gallazzi2014,vanderwel2008,sanchezblazquez2009,kovac2014,valentinuzzi2011,muzzin2012}. On
the other hand, spectroscopy of passive galaxies at $z$$\gtrsim$1.5
has been until recently very difficult: in spite of several
investigations pushing spectroscopic confirmation and more detailed
studies to higher redshifts
\citep[e.g.][]{cimatti2004,cimatti2008,daddi2005,kriek2006,kriek2009,onodera2012,vandesande2011,vandesande2013,toft2012,gobat2012,gobat2013,brammer2012,weiner2012,krogager2014,newman2014,
  belli2014}, sizable spectroscopic samples are still rare, and
studying $z$$>$1.5 passive populations mainly relies on photometric
samples
\citep[e.g.][]{wuyts2010,bell2012,ilbert2013,muzzin2013b,cassata2013}.
These studies show that the number density of passive galaxies rapidly
falls beyond $z$$>$1, so that by $z$$\sim$2 passive sources are no
longer the dominant population even among massive galaxies. However,
the observed evolution of massive cluster galaxies up to $z$$\sim$1
\citep[and also theoretical models, e.g.][]{delucia2006} typically
suggests early ($z$$\gtrsim$2-3) formation epochs for their stellar
populations \citep[e.g.,][]{mei2009,mancone2010,strazzullo2010}. We
might thus expect that the surge of passive galaxies around 10
billion years ago occurred differently in different environments.\\
\begin{figure}[hbp!]
\centering
\includegraphics[width=.48\textwidth,bb=85 370 540 700,clip]{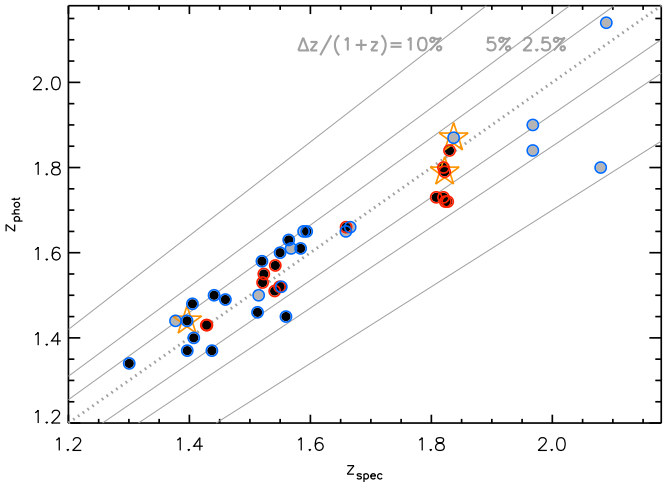}
\caption{Comparison of $z_{phot}$ vs. $z_{spec}$ for the full
  spectroscopic sample of $1.3<z<2.1$ passive galaxies. Lower quality
  $z_{spec}$ determinations are shown in gray. Blue and red circles
  mark sources from the VIMOS and O12 samples, respectively. Orange
  stars highlight MIPS-detected sources. Dotted and solid lines show
  the bisector and a relative scatter of 2.5, 5 and
  10\%. \label{fig:zszp}}
\end{figure}
\begin{figure*}[htl!]
\centering
\includegraphics[height=.15\textheight,bb=81 370 535 700,clip]{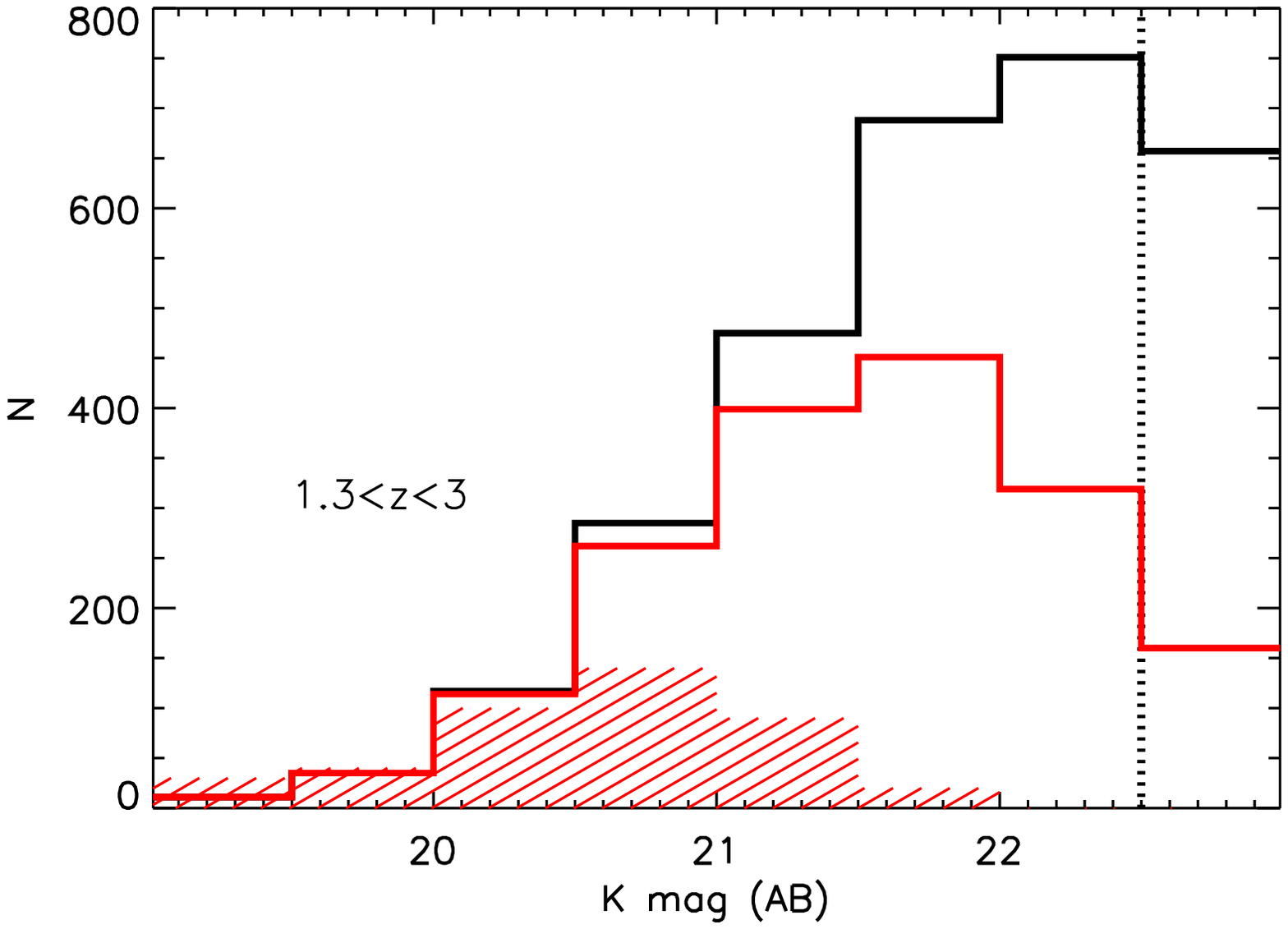}%
\includegraphics[height=.15\textheight,bb=135 370 535 700,clip]{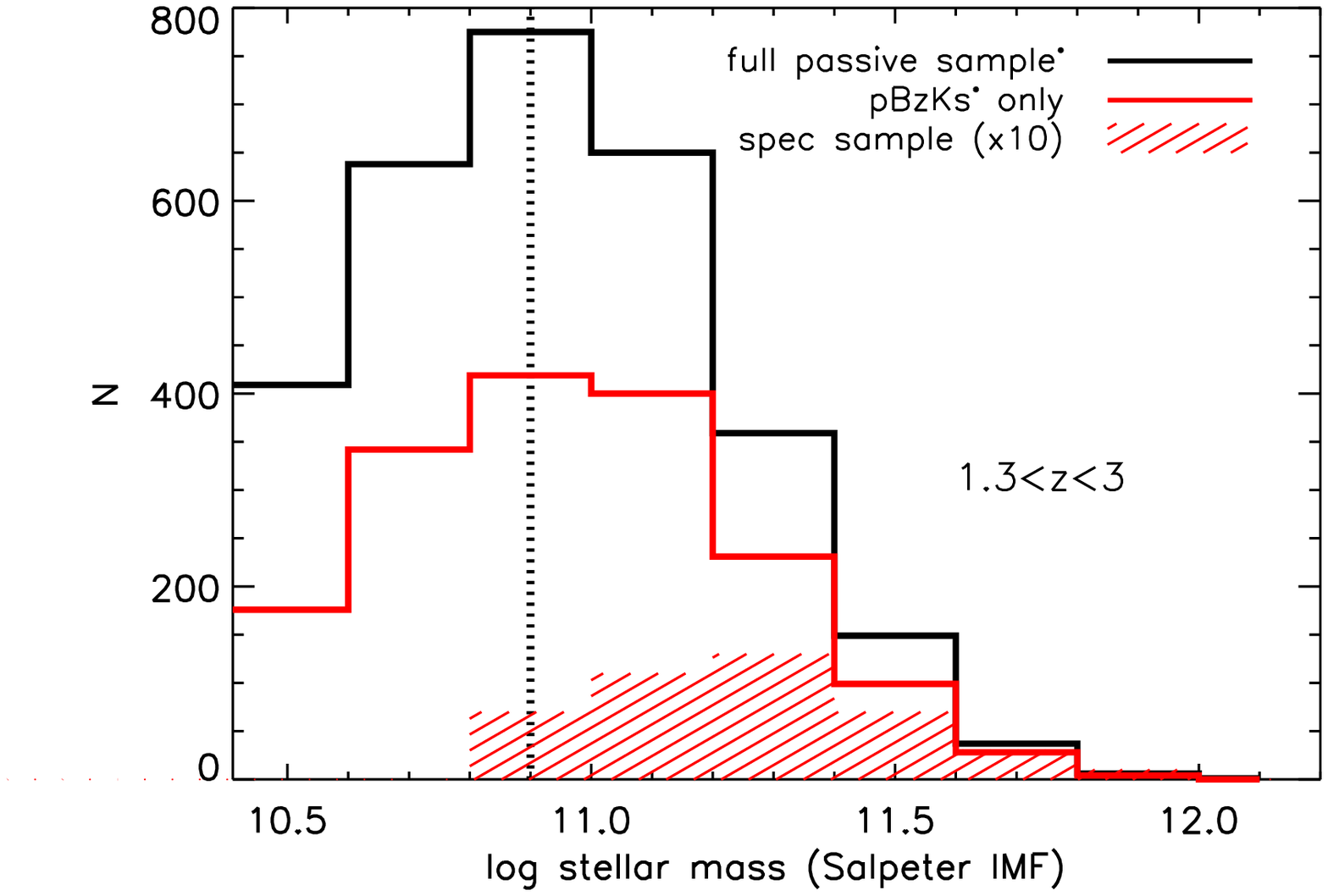}%
\includegraphics[height=.15\textheight,bb=135 370 535 700,clip]{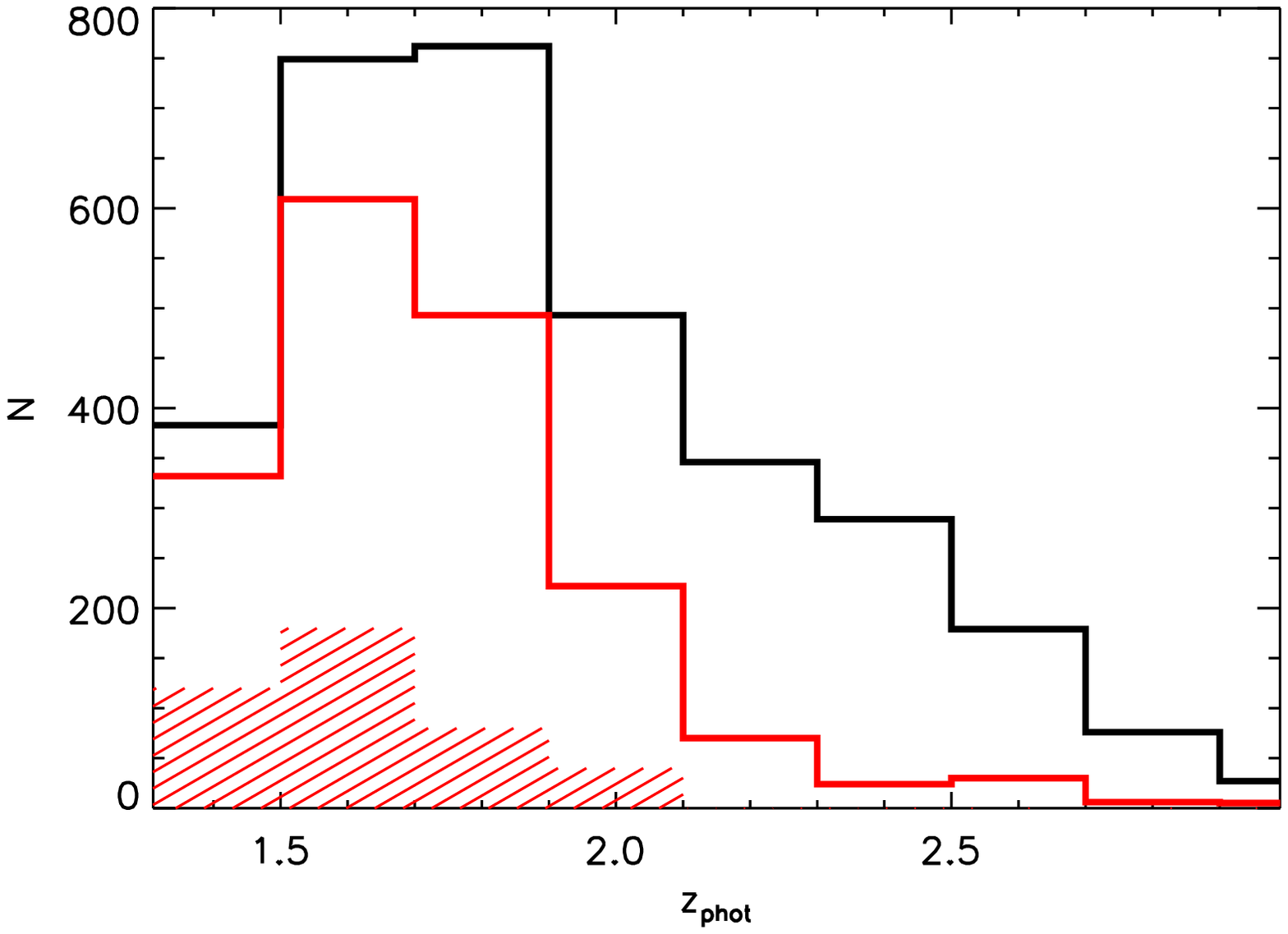}%
\includegraphics[height=.15\textheight,bb=132 367 535 701,clip]{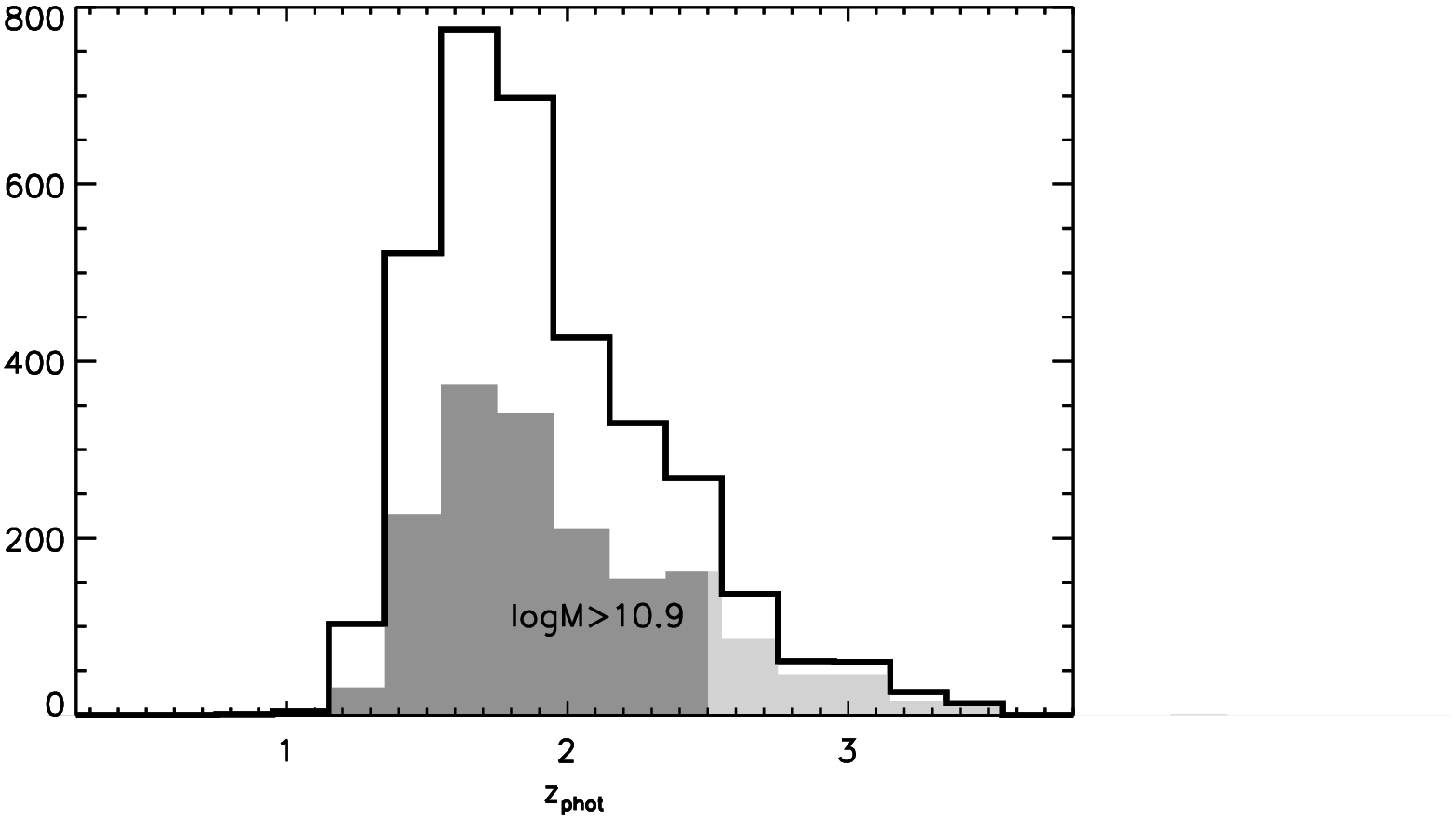}
\caption{Distributions of $K$-band magnitude, stellar mass and
  photometric redshift for the sample of $K_{AB}<23$, $z>1.3$ passive
  galaxies. The three leftmost panels show the full sample (see text,
  black line), its pBzK sub-sample (red line), and the spectroscopic
  sample (scaled by $\times 10$, red shaded area). Vertical dotted
  lines show the magnitude and mass (at $z<2.5$) completeness limits
  of the parent $K$-selected catalog (see text). The rightmost panel
  shows the full photo-z distribution of the $K_{AB}<23$ passive sample (black line), and
  of its log($M/M_{\odot}$)$>10.9$ sub-sample (shaded area,
  lighter-gray at $z>2.5$ where sample is beyond mass
  completeness). \label{fig:histos}}
\end{figure*}
Several recent studies have claimed evidence of significant star
formation even in central cluster regions at $z$$\gtrsim$1.5
\citep[among several
  others,][]{hilton2010,tran2010,hayashi2010,santos2011,fassbender2011,brodwin2013},
suggesting that indeed we are approaching the formation epoch of
massive cluster galaxies. However, it is also noticeable that massive
{\it passive} galaxies are often found even in such most distant
clusters, although in many cases sharing their environment with
galaxies in a still active formation phase \citep[e.g.,][]{kurk2009,
  papovich2010,
  gobat2011,gobat2013,tanaka2012,tanaka2013,spitler2012,strazzullo2013,newman2014,andreon2014}. This
may suggest that, even at a cosmic time when star formation rate
density is at its peak \citep[e.g.][]{madaudickinson2014}, and star formation
is still active in a considerable fraction of massive galaxies
\citep[e.g.][]{ilbert2013,muzzin2013b}, the densest cores of most evolved
cluster progenitors already host a typically small but characteristic
population of massive quiescent galaxies. For this reason,
overdensities of passive sources might be considered as possible
signposts to clusters at least up to $z$$\sim$2.

\section{Photometric redshift estimation for  high-redshift passive galaxies in the COSMOS field}
\label{sec:zspeczphot}

Using one of the first sizable samples of $z$$\gtrsim$1.4 passive
galaxies in the COSMOS field for calibration, in \citet{onodera2012}
(hereafter O12) we could estimate more accurate photometric redshifts
(photo--zs, $z_{phot}$) for high-redshift passive sources.  We have
now assembled a new, independent sample of passive galaxies in COSMOS
with redshifts measured through UV features using VLT/VIMOS
spectroscopy. We targeted 29 $I_{AB}$$<$25 galaxies selected as passive
BzKs \citep[``pBzKs'',][plus 6 24$\mu$m-detected pBzKs]{daddi2004} 
from the \citet{mccracken2010} (hereafter M10) catalog.  A redshift
was measured for 34 of the 35 targets, with a robust estimate
for 29 sources. The observations, analysis, and a full redshift list
will be presented in Gobat et al. (in prep.). Here we focus on a
sub-sample of 42 spectroscopically confirmed pBzKs, selected in the
range 1.3$<$$z_{spec}$$<$2.1 and with restframe UVJ colors
\citep{williams2009} consistent with passive populations (15 and 27
galaxies from the O12 and VIMOS samples, respectively, including 3
24$\mu$m-detected sources as noted below). Fig.~\ref{fig:zszp} shows
the performance on this sample of our photo-zs, estimated with EAzY
\citep{eazy} and calibrated as in O12. The normalized median absolute
deviation (NMAD) of $\Delta z /(1+z)$ is 2.5\% on the full sample, or
1.8\% excluding galaxies with less reliable $z_{spec}$
(Fig.~\ref{fig:zszp}), with no catastrophic outliers (thus $<$2.5\%
for this sample). All results presented here are based on the
photometric catalog by M10\footnote{An even better photo-z accuracy
  (as low as 1.5\%, with a marked improvement at $z$$\gtrsim$1.8) can
  be obtained calibrating photo-zs on this sample with the more recent
  photometry from the \citet{muzzin2013a} UltraVISTA catalog, in
  agreement with - and only marginally better than - the results
  obtained with \citet{muzzin2013a} photo-zs, as will be discussed in
  a forthcoming paper (Strazzullo et al., in prep.). Here we use the
  M10 catalog which includes the southern part of the
  2~sq.~deg. COSMOS field where one of our overdensities
  (Sec.~\ref{sec:overdensities}) is found, not covered by the
  UltraVISTA survey \citep{ultravistapaper}.}.
\begin{figure*}[htbp!]
\centering
\includegraphics[height=.21\textheight,bb=55 360 495 720,clip]{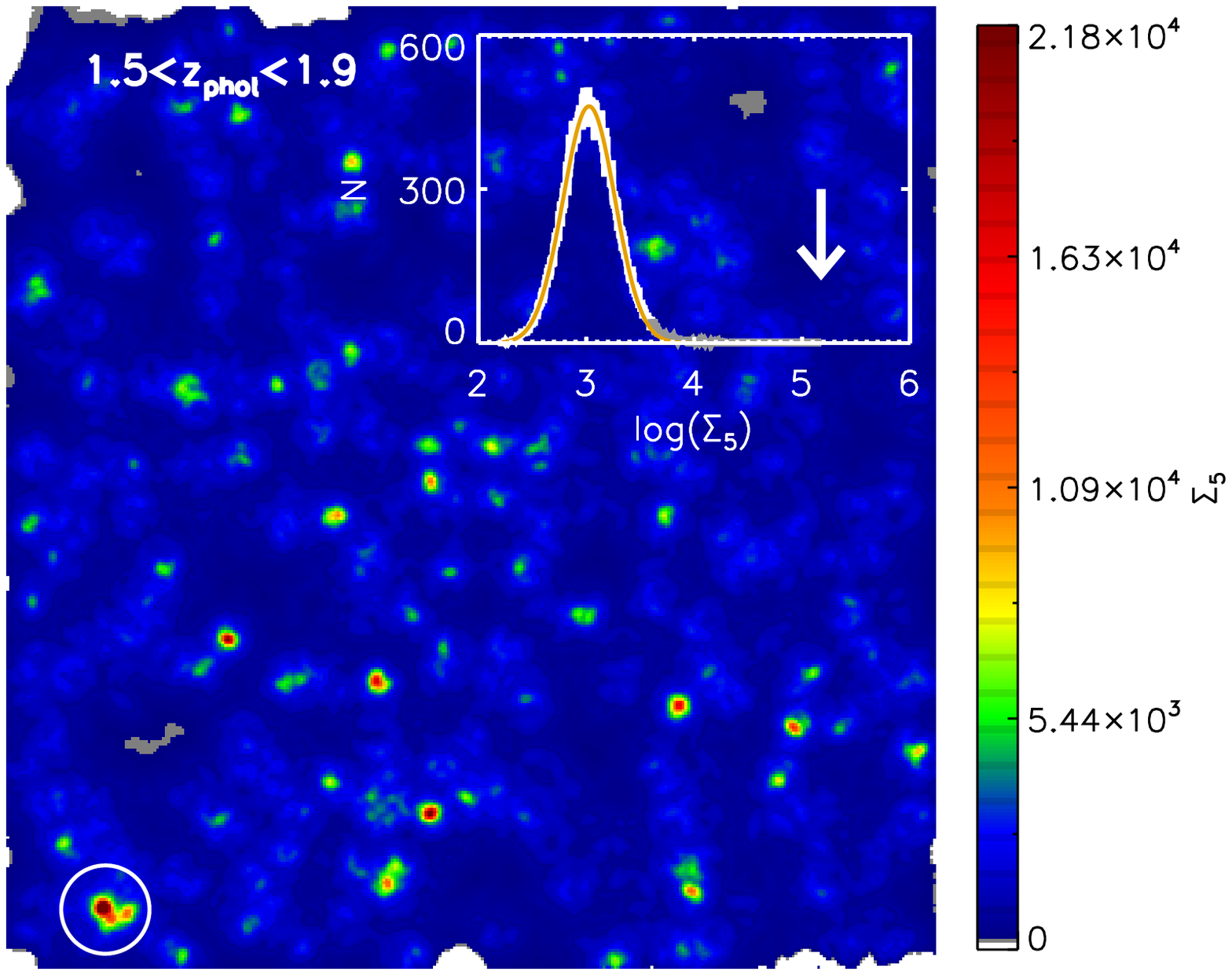}%
\includegraphics[height=.21\textheight,bb=55 360 495 720,clip]{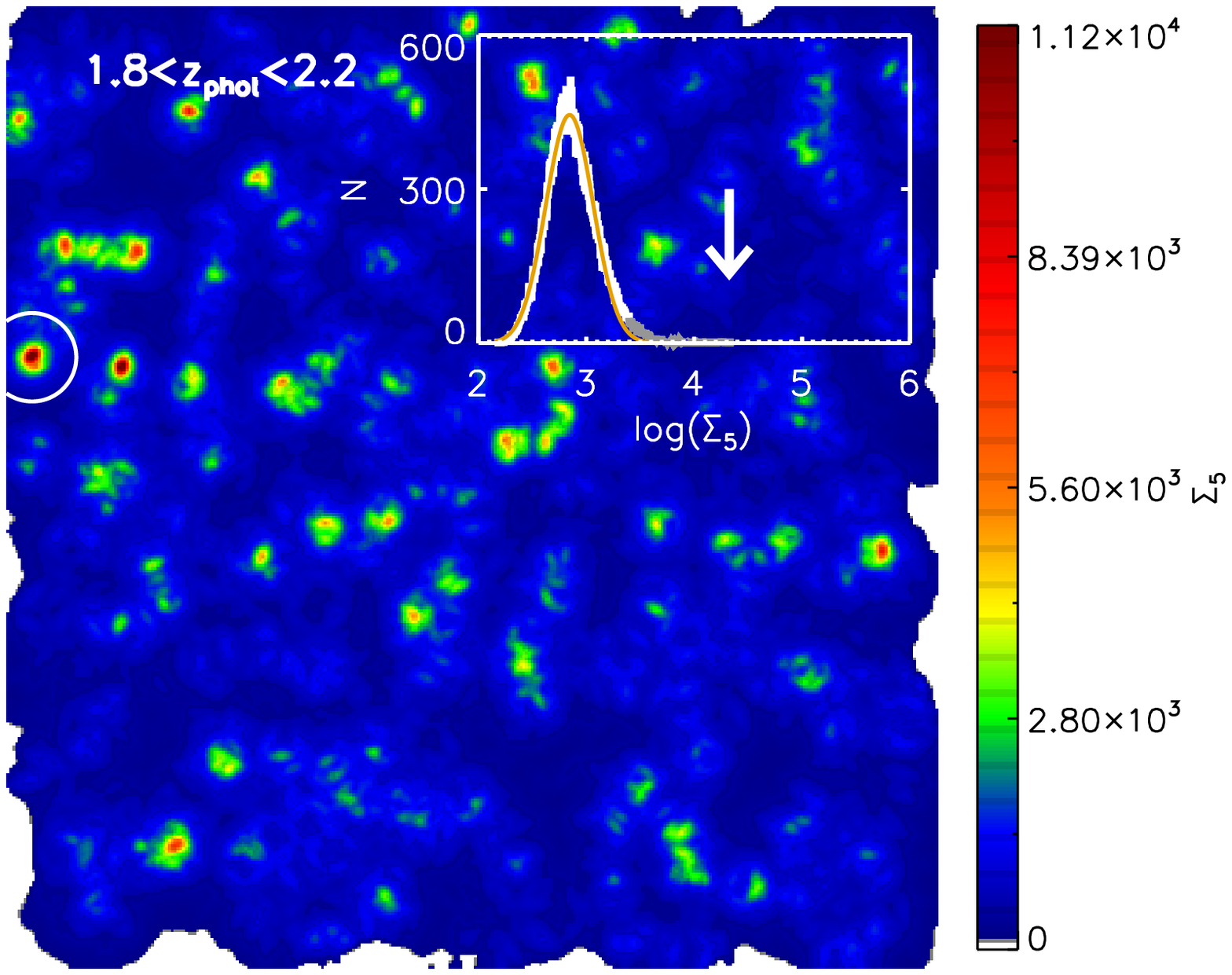}%
\includegraphics[height=.21\textheight,bb=55 360 487 720,clip]{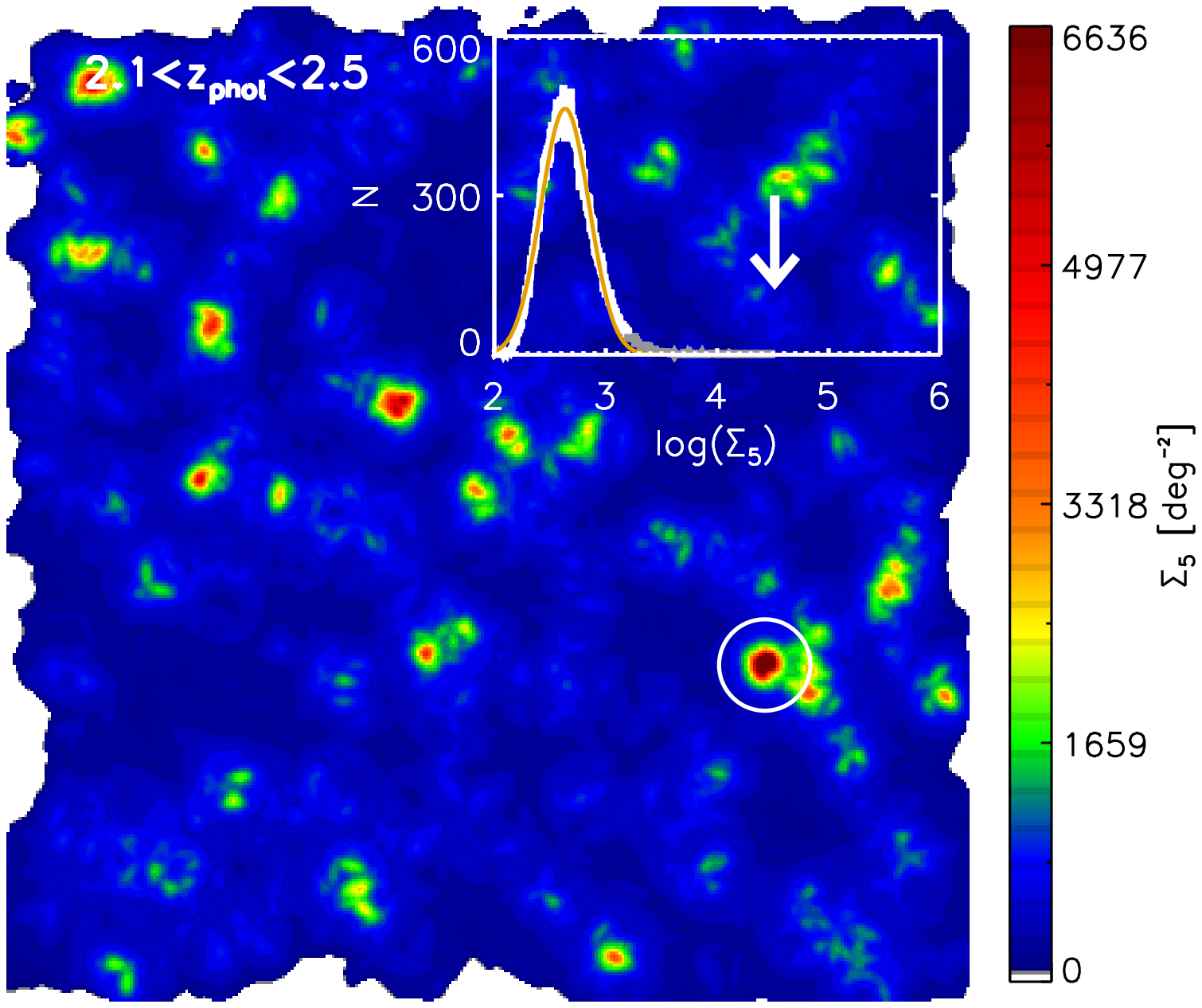}
\caption{Examples of $\Sigma_{5}$ maps in redshift slices of the full
  passive galaxy sample in the 2~sq.~deg. COSMOS field (North is up,
  East is left). Maps are scaled so that red colors correspond to
  $>5\sigma$ significance. The three candidate overdensities described
  in Sec.~\ref{sec:overdensities} are highlighted with white circles
  (2~Mpc radius proper). In each plot, the inset shows the
  distribution of log($\Sigma_{5}$) values in the map and its gaussian
  fit (white and orange lines, respectively), and the white arrow
  shows the peak $\Sigma_{5}$ value of the highlighted overdensity.
  \label{fig:sigma5maps}}
\end{figure*}
\section{Passive galaxy overdensities at $z>1.5$}
\label{sec:samplesel}
Spectroscopic confirmation of large passive galaxy samples at high
redshift is for now precluded, so we rely on passive candidates with
photo-zs calibrated as above. We select a sample of $z$$>$1.5 passive
galaxies as follows: with an initial BzK selection on the M10
$K_{AB}$$<$23 catalog, we take all pBzK galaxies, as well as galaxies
formally classified as star-forming BzKs (``sBzKs'') but having a
S/N$<$5 in the B- (and possibly z-) band. From this first selection,
we retain all galaxies that also have UVJ passive colors (assuming
their $z_{phot}$ as of Sec.~\ref{sec:zspeczphot}). Sources
detected at 24$\mu$m and satisfying the above criteria are retained,
because of the possibility of AGN-powered 24$\mu$m flux. Sources with
possibly contaminated IRAC photometry (as in the M10 catalog) were
discarded ($<$10\%).  In Fig.~\ref{fig:histos}, black and red lines
show the stellar mass, $K$-band magnitude and photo--z distributions
for the full retained sample of passive galaxies according to these
criteria, and for its subsample of pBzK sources, respectively.

 The $K_{AB}$$<$23 limit corresponds to a 90\% completeness for
 point-like sources, going down to $\sim$22.5 for disk-like profiles
 (see M10). At $z$$\sim$2.5 these limits would correspond to a mass
 completeness of log($M/M_{\odot}$)$\sim$10.8-11 ($\sim$10.9 in the
 following, \citet{salpeter1955} IMF) for an unreddened solar
 metallicity \citet{bc03} SSP formed at $z$$\sim$5. On the other hand,
 the combination of selection criteria adopted above is expected to
 result in a largely pure (in terms of contaminants) but {\it not
   complete} sample of massive (log($M/M_{\odot}$)$\gtrsim$10.9)
 passive galaxies. For this reason, we may be missing some
 overdensities or reducing their significance, which would affect in a
 conservative way the results presented here.  The BzK selection, and
 the depth of the M10 catalog, effectively limit our sample at
 $z$$>$1.5 and $z$$<$2.5, respectively (Fig.~\ref{fig:histos}, right).
 The full sample of $K_{AB}$$<$23 passive galaxy candidates includes
 $\sim$4500 sources. Of these, $\sim$3500 are at
 1.5$\leq$$z_{phot}$$\leq$2.5 ($\sim$70\% at
 log(M/M$_{\odot}$)$>$10.9, $\sim$50\% selected as pBzK, $\sim$10\%
 24$\mu$m-detected).
\subsection{Identification of cluster candidates}

\label{sec:overdensities}
\begin{figure*}[htbl!]
\centering
\includegraphics[height=.195\textheight,bb=83 364 542 704,clip]{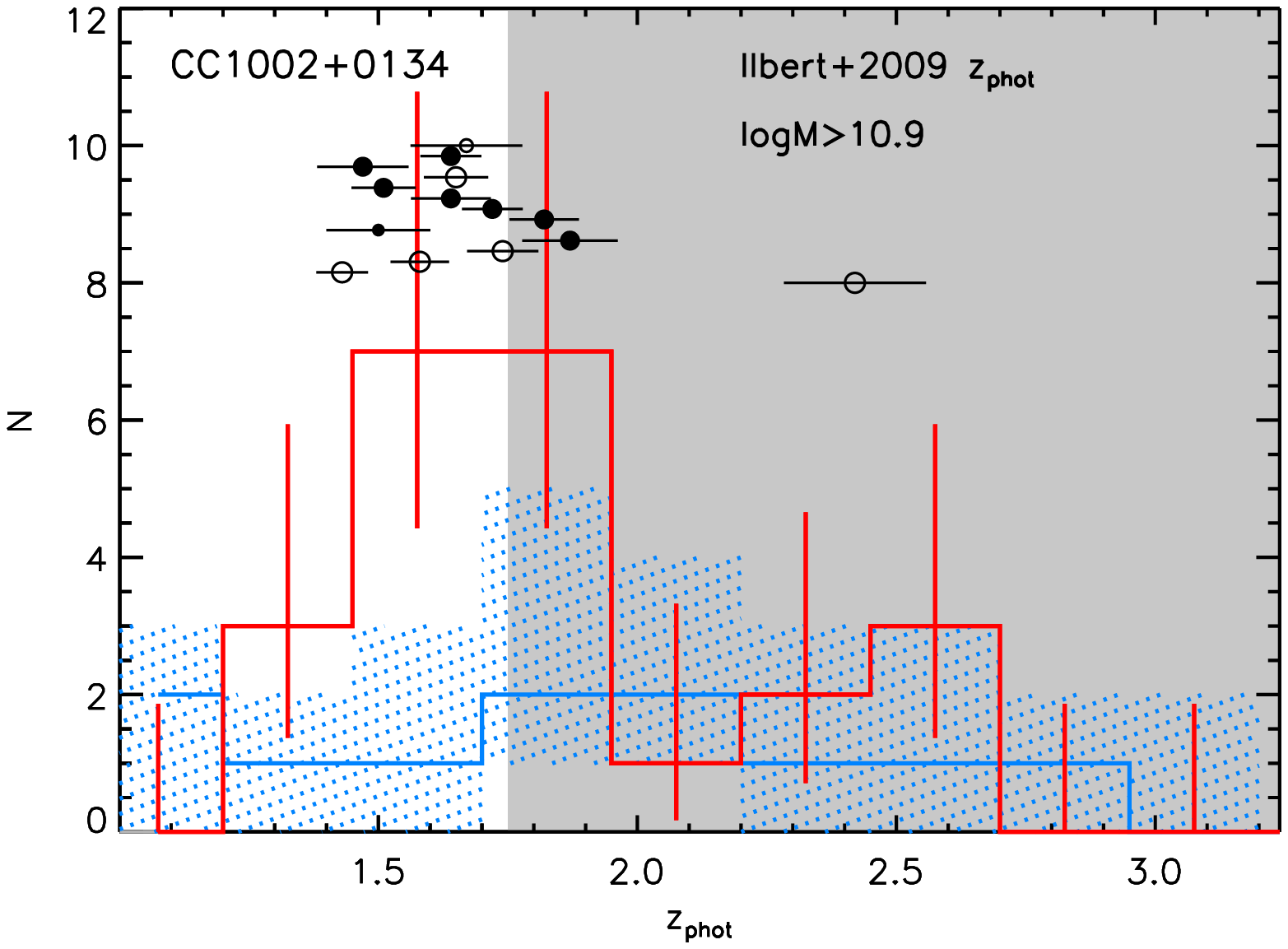}%
\includegraphics[height=.195\textheight,bb=127 364 542 704 ,clip]{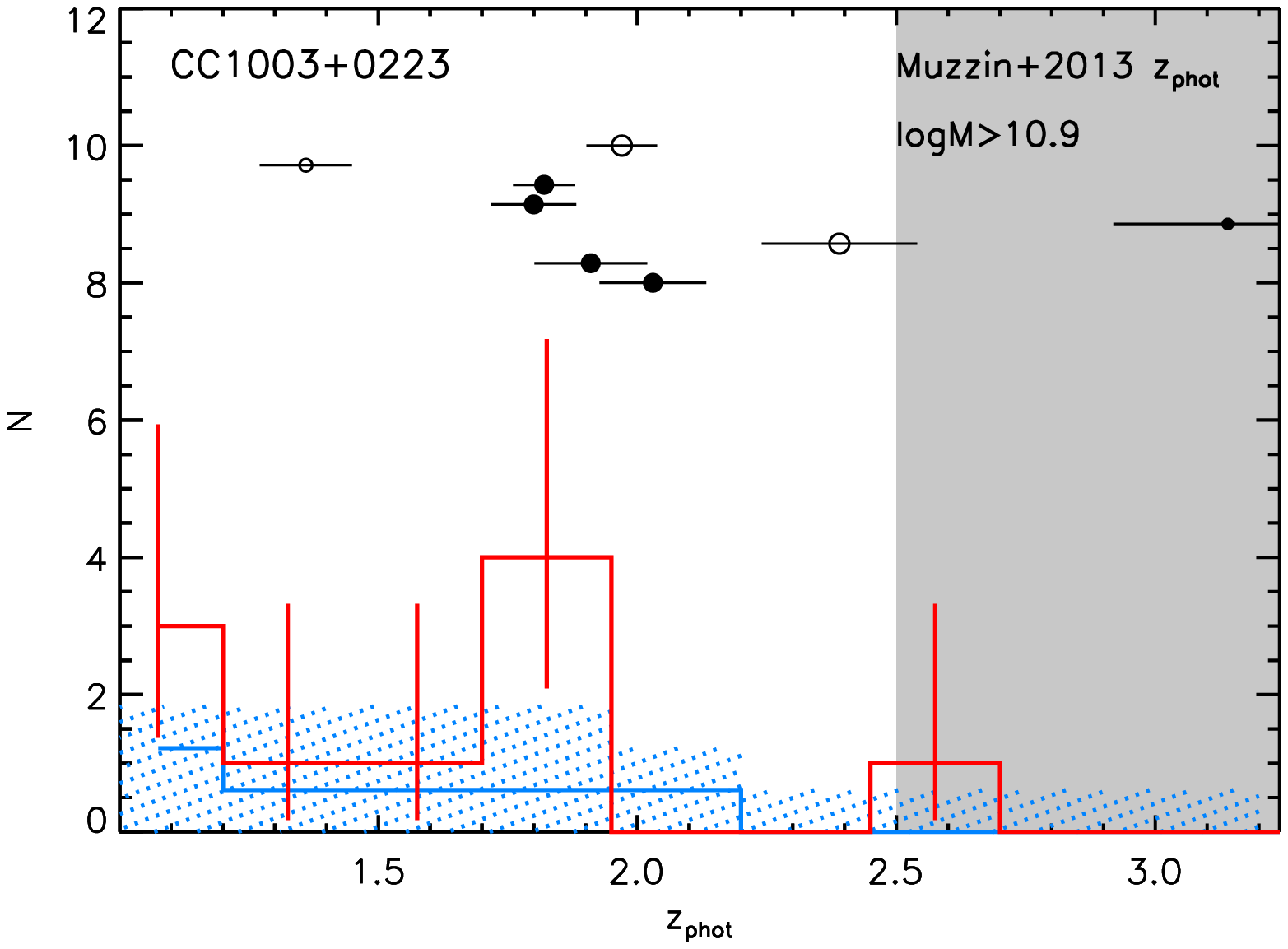}%
\includegraphics[height=.195\textheight,bb=127 364 542 704,clip]{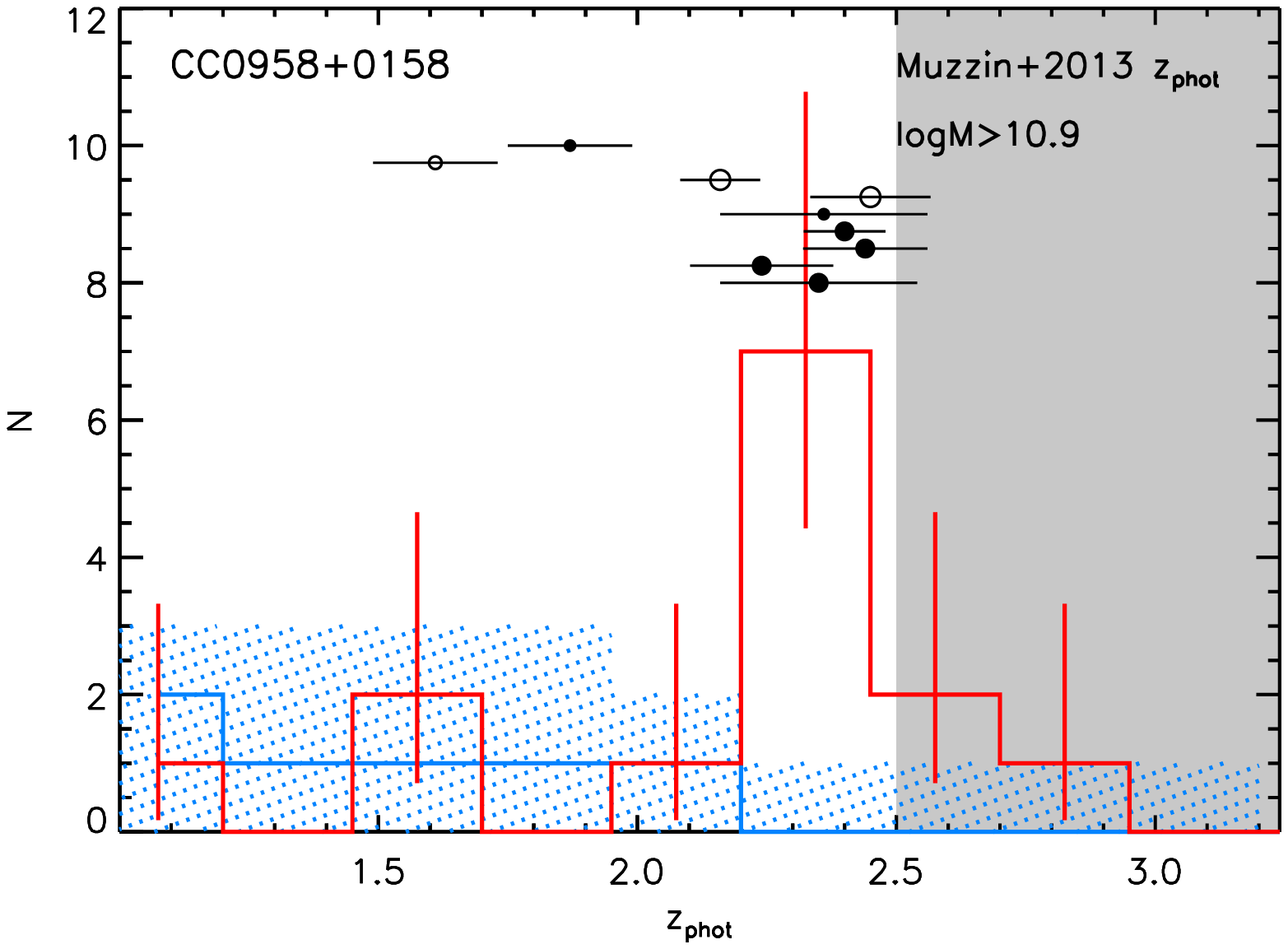}
\caption{Photo--z distributions for the {\it whole} (passive and
  star-forming) population of log(M/M$_{\odot}$)$>10.9$ galaxies
  within an aperture of $r<600$~kpc from the three passive
  overdensities as labeled \citep[red lines, errors as from][]{gehrels86}.  Galaxies and photo--zs used here are from 
  \citet{muzzin2013a}  for CC1003+0223 and CC0958+0158, and
  from \citet{ilbert2009} for CC1002+0134 (see text). In each panel,
  the blue line and dotted area show the median and $16^{th}-84^{th}$
  percentiles of the photo--z distribution (from the same catalogs) in
  $r<$600kpc apertures at 100 random positions in the COSMOS field,
  while the grayed-out region shows the redshift range where the
  sample is no longer mass-complete. Black filled (empty) circles
  scattered above the histograms (at random y-axis coordinates) show
  {\it our photo-z determinations for the passive sample used in this
    work} within 300~kpc (600~kpc) from the overdensity center
  (larger/smaller symbols show galaxies more/less massive than
  log(M/M$_{\odot}$)=10.9).
  \label{fig:zphotdensplot}}
\end{figure*}
We build local density maps for the full COSMOS field in redshift
slices, based on the catalog of passive galaxy candidates described
above. In spite of the quoted $\sim$2\% photo-z relative accuracy (by
comparison with spectroscopic redshifts, Sec.~\ref{sec:zspeczphot}),
we need to consider that the spectroscopic sample is very biased
towards brighter, lower-redshift sources (Fig.~\ref{fig:histos}), thus
photo-z performance on the bulk of our sample is likely significantly
worse. We estimated a more realistic photo-z accuracy as a function of
magnitude, recalculating photo-zs on SEDs of well-fitted spectroscopic
sources ($|\Delta z|$/(1+z)$<2.5\%$) dimmed to fainter magnitudes,
randomly scattering fluxes in the different bands according to
photometric errors in our catalog. This simulation indeed gives a
photo-z accuracy $<$2.5\% at $K_{AB}$$\lesssim$21
(Sec.~\ref{sec:zspeczphot}), but rising to $\sim$3.5\%~(5\%) at
$K_{AB}$$\sim$22~(22.5), and to more than 6\% approaching our
$K_{AB}\sim$23 limit. These estimates are ``model-independent'' in the
sense that they use observed (rather than synthetic) SEDs, but they
still assume that SEDs of less massive and/or higher redshift sources
in our sample behave similarly to those of the spectroscopic sources
used as inputs in the simulation.  For comparison, the formal 68\%
errors estimated by EAzY on the simulated SEDs would be on
average 40-50\% larger (reaching $\sim$8\% at $K_{AB}$$\sim$23). For
this reason, we build our density maps in redshift slices of
$\Delta$$z$=$\pm$0.2 (corresponding to a $\pm$1$\sigma$ relative
accuracy of 6-8\% at 1.5$<$$z$$<$2.5) with a step of 0.05 in central
redshift.

We use the $\Sigma_{5}$ (5$^\mathrm{th}$ nearest neighbour) density
estimator, which for our sample probes a median (over the full map)
distance of $\sim$1.4~Mpc in the $z$$\sim$2 slice, with minimum and
maximum distances of 100-200~kpc and 4-5~Mpc in all redshift slices,
thus properly probing the typical scales we are investigating.  For
comparison, a $\Sigma_{3}$ estimator would also probe such scales
(median distance $\sim$1.1~Mpc at $z$$\sim$2, minimum/maximum
distances $\sim$50~kpc and 3-4~Mpc), while $\Sigma_{7}$ would probe
median distances closer to 2~Mpc with minimum/maximum of 0.5/4.5~Mpc
at $z$$\sim$2, thus becoming less sensitive to the scales we need to
probe. Fig.~\ref{fig:sigma5maps} shows three examples of $\Sigma_{5}$
maps.  For each map, we estimate the significance of overdensities by
fitting the distribution of log($\Sigma_{5}$) in the whole map with a
Gaussian (Fig.~\ref{fig:sigma5maps}).

 At the same time, we also use
an independent approach to search for concentrations of massive
passive galaxies with consistent photo--zs in a very small
cluster-core sized area.  In particular, based on observations of the
$z=2$ cluster Cl~J1449+0857 \citep[][other examples in
  Sec.~\ref{sec:intro}]{gobat2011,gobat2013,strazzullo2013}, we search
our log(M/M$_{\odot}$)$>$10.9 passive sample for sources with at least
three other passive galaxies within $|\Delta z|$/(1+$z$)$<$7.5\%
(accounting for photo-z uncertainties described above) and a physical
distance $\leq150$~kpc\footnote{Based on our sample of
  log(M/M$_{\odot}$)$>$10.9 passive galaxies, the Poissonian
  probability of finding $\geq3$ neighbours at $\leq150$~kpc (or
  $\geq4$ sources within a radius of $\leq150$~kpc) with a $|\Delta
  z|$/(1+$z$)$<7.5\%$ is $<5\times10^{-4}$ ($<2\times10^{-7}$,
  respectively) at all redshifts probed.}. This approach independently
retrieves three most significant ($\gtrsim 7 \sigma$ in the
$\Sigma_{5}$ - as well as $\Sigma_{3}$ - maps) overdensities
highlighted in Fig.~\ref{fig:sigma5maps}, and described here
below:

 $\bullet$ CC1002+0134 at 1.5$\lesssim$$z_{phot}$$\lesssim$ 1.8 -- A
concentration of passive sources around RA, Dec $\sim$$10^h02^m40^s$,
$+01\degree34^m20^s$, with 4 galaxies within a radius $r=90$~kpc and
$z_{phot}$ within $\leq$1.5$\sigma$ (given each source magnitude,
based on the simulation described above) from a mean
$z_{phot}$$\sim$1.76. However, the photo-z distribution of the central
sources in this candidate overdensity is quite broad compared to the
expected photo-z uncertainties (see
Fig.~\ref{fig:zphotdensplot}). This might suggest a chance
superposition of passive galaxies, or possibly of different, unrelated
structures, along the line of sight at $1.5<z<1.8$. 
On the other hand, \citet{aravena2012} already reported the
identification of a candidate cluster at $z_{phot}\sim1.55$ at the
same position, based on an overdensity of galaxies with
1.5$<$$z_{phot}$$<$1.6, a radio source at a consistent redshift, a
tentative detection of extended X-ray emission, and the presence of a
small population of passive sources. The actual nature of this
structure is thus still unclear.

 $\bullet$ CC1003+0223 at $z_{phot}$$\sim$1.90 -- A concentration of 4
passive galaxies at RA, Dec$\sim$$10^h03^m05^s$,$+02\degree23^m24^s$
within $r$$<$110~kpc and $z_{phot}$ within $\lesssim$1.4$\sigma$ from
the mean $z_{phot}$ (a further galaxy with $z_{phot}$ consistent
within 1$\sigma$ is found at $r$$<$340~kpc). Half of these 4 sources
were selected as UVJ-passive sBzKs (see Sec.~\ref{sec:samplesel}). None is
24$\mu$m detected. All are consistent with being UVJ-passive also in
the \citet{muzzin2013a} catalog, three out of four with a photo-z
consistent within 1$\sigma$ with the mean $z_{phot}$ estimated here.
The redshift distribution of the central sources is consistent with
the presence of a single structure. Another overdensity of similar
significance at a consistent photo-z is visible in the $\Sigma_5$ map at
$<4$~Mpc West (Fig.~\ref{fig:sigma5maps}). 
   
 $\bullet$ CC0958+0158 at $z_{phot}$$\sim$2.35 - A concentration of 4
passive galaxies at RA, Dec$\sim$$9^h58^m53^s$,$+01\degree58^m01^s$
within $r$$<$130~kpc and a $z_{phot}$ within $\lesssim$0.8$\sigma$
from the mean $z_{phot}$ (one further, lower-mass galaxy at the same
$z_{phot}$ is found at $r$$<$290~kpc). Given their redshift (thus
faintness) all of these sources were selected as UVJ-passive sBzKs
(see Sec.~\ref{sec:samplesel}). One might be associated with a
24$\mu$m detection. All are consistent with being UVJ-passive also in
the \citet{muzzin2013a} catalog, with photo-zs consistent within
1$\sigma$ with the mean $z_{phot}$ estimated here. The photo-z
distribution is very compact, consistent with a single structure. An
overdensity at a consistent position and redshift is also
visible in \citet{scoville2013} density maps.  We also note the
proximity of the \citet{spitler2012} cluster candidate at similar
redshift, $\sim$14~Mpc NE.  \citet{chiang2014} also claim the presence
of several proto-cluster candidates within a few Mpc of CC0958+0158 at
a similar photo-z.

 In Fig.~\ref{fig:zphotdensplot} we show for
comparison the photo--z distribution of a mass-limited sample of the
{\it whole} (passive and star-forming) galaxy population in the
surroundings ($r<600$~kpc) of each passive overdensity, with respect
to the distribution in same-size apertures at 100 random positions in
the COSMOS field. These distributions in Fig.~\ref{fig:zphotdensplot}
are based on public galaxy catalogs and photo-z determinations,
totally independent from those we use in this work. For
CC1003+0223\footnote{This overdensity is at the edge of the area
  probed by our catalog (see Fig.~\ref{fig:sigma5maps}). For this
  reason, the $r<$600~kpc aperture is not fully covered by the
  catalog; the field distribution in Fig.~\ref{fig:zphotdensplot} is
  thus scaled accordingly, by the effectively covered area. } and
CC0958+0158 we use the \citet{muzzin2013a} UltraVISTA catalog, while
for CC1002+0134 - not covered by the UltraVISTA survey - we use the
\citet{ilbert2009} catalog (note that this is an $i$-selected catalog,
thus not optimal in this redshift range, as shown by the gray region
in Fig.~\ref{fig:zphotdensplot}, left).  We also show in
Fig.~\ref{fig:zphotdensplot} {\it our} photo-z determinations for {\it
  our passive galaxy sample} around the overdensities in the same
aperture and mass range. Although clearly affected by limited
statistics, Fig.~\ref{fig:zphotdensplot} shows the correspondence
between {\it our} photo-zs of {\it passive} sources identifying the
overdensities, and the excess in the redshift distribution from {\it
  completely independent} photo-z determinations of the {\it whole}
galaxy population in their surroundings.
\section{Discussion and summary}
This letter investigates the possible identification of first
cluster-like environments using evolved galaxy populations as tracers
of an early-quenched cluster core. We described the identification of
3 candidate overdensities of passive galaxies, selected in
redshift-sliced density maps and with properties similar to passive
galaxy concentrations in $z$$\sim$2 clusters. This study relies on
accurate photo-z determination for high-redshift passive sources
calibrated on one of the largest spectroscopic samples available to
date. We present here only the first results on the strongest
candidate overdensities. Further investigation focusing on alternative
sample selections and the identification of lower-mass structures,
with improved photo-zs based on more recent, deeper photometry (Sec.~
\ref{sec:zspeczphot}), will be presented in a forthcoming paper.

 We currently have no proof that the candidate overdensities we
 identified are real structures. Even if they were actually clusters,
  their expected mass and redshift would put them beyond reach of the
 Chandra C-COSMOS \citep{elvis2009} and XMM-Newton
 \citep{hasinger2007,cappelluti2009} programs in COSMOS, which place
 3$\sigma$ limits of 4-9$\times10^{43}$~erg~s$^{-1}$ on their X-ray
 luminosity, thus 5-7$\times 10^{13}$M$_{\odot}$ on their mass
 \citep{leauthaud2010}. This would be consistent with their
 similarities with the passive concentration in Cl J1449+0857 \citep[
   M$\sim5\times10^{13}$~M$_{\odot}$,][]{
   gobat2011,gobat2013,strazzullo2013}. As a reference, in a WMAP7
 \citep{komatsu2011} cosmology we expect to find $\sim$2-8 structures
 more massive than 5-7$\times$$10^{13}$~M$_{\odot}$ in the
 1.7$<$$z$$<$2.5 range in a 2 sq.~deg. field \citep[or a factor
   $\sim$2 higher with a Planck cosmology,][]{planck.XVI.2014}.

\begin{figure}[t!]
\centering
\includegraphics[width=.4\textwidth,bb=107 370 532 693,clip]{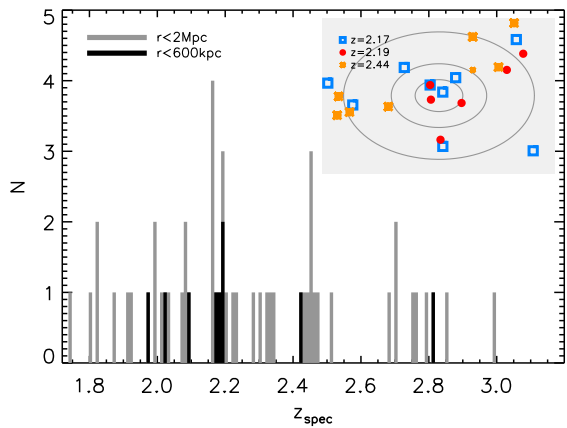}
\caption{The distribution of spectroscopic redshifts around the CC0958+0158 overdensity (see
text). The gray and
  black histograms show the distribution within 600 and 2000~kpc,
  respectively. The inset shows the spatial distribution around the
  overdensity center of galaxies within $\Delta z
  \sim0.1$ (or $\Delta z \sim 0.2$, smaller symbols) of the three
  spikes at $z\sim$2.17, 2.19 and 2.44, as indicated. The three gray circles
  have radii of 0.5, 1, 2~Mpc proper.
  \label{fig:zspecdist2.3}}
\end{figure}

 A final confirmation necessarily relies on spectroscopic follow-up,
 which is not available yet. For the time being, just for the most
 distant of our candidate overdensities, covered by the outer part of
 the zCOSMOS-deep survey \citep[][ and in prep.]{lilly2007} and by a
 Subaru/MOIRCS spectroscopy program \citep{valentino2015}, we
 have been able to combine available spectroscopic redshifts -- all of
 star-forming galaxies -- to tentatively probe the redshift
 distribution in its surroundings. This small, {\it
   a-posteriori}-assembled spectroscopic sample, is hampered by poor
 sampling of the central, densest cluster-candidate region, and
 sub-optimal target selection. Nonetheless, a possible redshift spike
 appears at $z\sim2.19$ with 5(3) galaxies at $2.18 \lesssim z_{spec}
 \lesssim 2.2$ within $\sim$1600~(500)~kpc from the overdensity
 center, plus two more spikes of galaxies within 2~Mpc of the center,
 at $z\sim2.17$ and $z\sim2.44$ (6 and 5 galaxies within $\Delta z
 \sim \pm 0.01$, respectively, Fig.~\ref{fig:zspecdist2.3}). Although
 the spikes contain a similar number of galaxies, their spatial
 distribution (Fig.~\ref{fig:zspecdist2.3}) might suggest that the
 passive overdensity is more likely at $z\sim2.2$. Even if somewhat
 lower than the estimated photo-z of the passive galaxies, this would
 be anyway consistent within the uncertainties.

Dedicated follow-up work is obviously still needed to verify whether
these candidate overdensities are indeed signposts for early cluster
environments. If successful, this approach would provide a further
option to extend the investigation of distant cluster-like structures
to the $z\sim2-2.5$ range. In comparison with most other
(proto-)cluster search techniques at these redshifts, e.g. the ``IRAC
selection'' \citep[e.g., ][]{papovich2010,stanford2012}, targeted
searches around radio-galaxies
\citep[e.g.,][]{venemans2007,wylezalek2013}, 3D mapping \citep[ with
  spectroscopic or photometric redshifts, e.g.,
][]{diener2013,scoville2013,chiang2014,mei2014}, or -- closer to what
done here -- overdensities of optically red galaxies
\citep[e.g.,][]{andreon2009,spitler2012}, the approach discussed in
this work is likely to favour, by definition, the most evolved
environments, allowing a better probe of the diversity of cluster
progenitors at a crucial time for the formation of both clusters and
their massive galaxies.

\begin{acknowledgements}
  We thank M. Pannella and A. Saro for helpful inputs. VS, ED, RG and
  FV were supported by grants ERC-StG UPGAL 240039 and
  ANR-08-JCJC-0008. AC and MM acknowledge grants ASI n.I/023/12/0 and
  MIUR PRIN 2010-2011  "The dark Universe and the cosmic evolution of
  baryons; from current surveys to Euclid''. Based on observations from ESO Telescopes
  under program IDs 086.A-0681, 088.A-0671, LP175.A-0839, and
  179.A-2005.
\end{acknowledgements}


\bibliographystyle{aa}
\bibliography{refs}

\makeatletter
\if@referee
\processdelayedfloats
\pagestyle{plain}
\fi
\makeatother
\end{document}